\documentclass[12pt]{article}
\usepackage{amssymb}
\usepackage{graphicx}
\usepackage{amsmath}

\input{tcilatex}
\input{tcilatex}
\begin{document}

\begin{center}
\bigskip {\Large Uniqueness of two phaseless inverse acoustics problems in
3-d }

Michael V. Klibanov

Department of Mathematics and Statistics

\ \ \ \ \ \ \ \ \ \ \ University of North Carolina at Charlotte

\ \ \ \ \ \ \ \ \ \ \ Charlotte, NC 28223, U.S.A.

\ mklibanv@uncc.edu{\Large \ }

\bigskip

\textbf{Abstract}
\end{center}

Uniqueness is proven for two 3-d inverse problems of the determination of
the spatially distributed sound speed in the frequency dependent acoustic
PDE. The main new point is the assumption that only the modulus of the
scattered complex valued wave field is measured on a certain set.

\section{Introduction}

\label{sec:1}

When considering Coefficient Inverse Problems (CIPs) in the frequency
domain, it is usually assumed that both modulus and phase of the complex
valued function representing the wave field is known on a certain set, see,
e.g. \cite{Nov1,Nov2}\ for global uniqueness results and reconstruction
methods. However, it is impossible to measure the phase in many
applications. In these applications only the modulus of the scattered
complex valued wave field can be measured (section 1.3). Therefore, it is
worthy to investigate CIPs in the frequency domain, assuming that only the
modulus of the scattered wave field is known on a certain set.

In the recent work \cite{Kr} the author has proven uniqueness theorems for
four inverse scattering problems of determining the compactly supported
potential $q\left( x\right) ,x\in \mathbb{R}^{3}$ in the Schr\"{o}dinger
equation in the case when only the modulus of the complex valued wave field
is measured on a certain set and the phase is unknown. The goal of the
current publication is to extend the result of \cite{Kr} to the case of the
3-d acoustic equation with the unknown spatially varying sound speed. The
author is unaware about previous similar results for the acoustic equation
in $n-$d, $n=1,2,3$.

Below $C^{s+\alpha }$ are H\"{o}lder spaces, where $s\geq 0$ is an integer
and $\alpha \in \left( 0,1\right) .$ Let $\Omega ,G\subset \mathbb{R}^{3}$
be two bounded domains. Let $G_{1}\subset \mathbb{R}^{3}$ be a convex
bounded domain with its boundary $S\in C^{1}$.\ Let $\varepsilon \in \left(
0,1\right) $ be a number. Below 
\begin{equation}
\Omega \subset G_{1}\subset G,S\cap \partial G=\varnothing .  \label{1}
\end{equation}%
\begin{equation}
dist\left( S,\partial \Omega \right) >\varepsilon ,  \label{2}
\end{equation}%
where $dist\left( S,\partial \Omega \right) $ is the Hausdorff distance
between $S$ and $\partial \Omega .$ Let $c\left( x\right) $ be the variable
sound speed satisfying the following conditions 
\begin{equation}
c\in C^{5}\left( \mathbb{R}^{3}\right) ,c\left( x\right) =1\text{ for }x\in 
\mathbb{R}^{3}\diagdown G,  \label{1.1}
\end{equation}%
\begin{equation}
c\left( x\right) \geq c_{0}=const.>0,\forall x\in G.  \label{1.2}
\end{equation}%
In addition, we assume that there exists a point $x_{0}\in \Omega $ such
that 
\begin{equation}
\left( x-x_{0},\nabla c^{-2}\left( x\right) \right) \geq 0,\forall x\in 
\overline{G}.  \label{1.3}
\end{equation}

Note that usually the minimal smoothness of unknown coefficients is of a
minor concern of uniqueness theorems for multidimensional CIPs, see, e.g. 
\cite{Nov1,Nov2} and Theorem 4.1 in \cite{Rom}. The $C^{5}-$smoothness
condition of $c\left( x\right) $ is imposed because we need to use Theorem
3.1 of \cite{Kl16}. This theorem, in turn requires the $C^{4}-$smoothness of
the solution of the Cauchy problem for the acoustic equation in the time
domain, see section 2 for this problem. To establish that $C^{4}-$%
smoothness, we refer in section 2 to Theorem 2.2 of \cite{Ktherm}, which
requires $c\in C^{5}\left( \mathbb{R}^{3}\right) $. The survey \cite{Kl16}
is about the method of proofs of global uniqueness and stability theorems
for multidimensional non over-determined CIPs for PDEs, which was originally
proposed in \cite{BukhKlib}. This method is based on Carleman estimates.

\textbf{Lemma 1}. \emph{Assume that conditions (\ref{1.1})-(\ref{1.3}) are
in place. Then the family of geodesic lines generated by the function }$%
c\left( x\right) $\emph{\ holds the non-trapping property in} $\mathbb{R}%
^{3}.$

\textbf{Proof}. The validity of this lemma follows immediately from formulae
(3.23') and (3.24) of section 2 of chapter 3 of the book \cite{Rom}. $%
\square $

Consider the function $g\left( x\right) $ satisfying the following
conditions 
\begin{equation}
g\in C^{7}\left( \mathbb{R}^{3}\right) ,g\left( x\right) =0\text{ in }%
\mathbb{R}^{3}\diagdown G,  \label{1.10}
\end{equation}%
\begin{equation}
g\left( x\right) \neq 0,x\in S.  \label{1.11}
\end{equation}%
It would be probably better to assume in Inverse Problems 1,2 below that $%
g\left( x\right) =\delta \left( x-x_{0}\right) $ for a certain source
position $x_{0}\in \mathbb{R}^{3}.$ However, even if the entire wave field,
rather than only its modulus, would be measured, still uniqueness theorems
for corresponding CIPs for the 3-d acoustic equation in the case $g\left(
x\right) =\delta \left( x-x_{0}\right) $ are currently known only if the
data are over-determined ones, see, e.g. \cite{Kok,Lavr}.\ This is the case
of infinitely many measurements when the number of free variables in the
data exceeds the number of free variables in the unknown coefficient. The
above mentioned technique of \cite{BukhKlib,Kl16} is currently the only one,
which enables to prove global uniqueness for multidimensional CIPs with the
data resulting from a single measurement event. The data in this case are
non over-determined ones. On the other hand, this technique requires that%
\begin{equation}
\Delta g\left( x\right) \neq 0,\forall x\in \overline{G}_{1}.  \label{1.141}
\end{equation}

To mitigate the concern about $\delta \left( x-x_{0}\right) $, consider an
analog of examples of \cite{Kl16,Kr}. Let the function $\chi \left( x\right)
\in C^{\infty }\left( \mathbb{R}^{3}\right) $ be such that $\chi \left(
x\right) =1$ in $G_{1}$ and $\chi \left( x\right) =0$ for $x\notin G.$ The
existence of such functions $\chi \left( x\right) $ is well known from the
Real Analysis course. Let the point $x_{0}\in \overline{G}_{1}.$ For a
number $\sigma >0$ consider the function $\delta _{\sigma }\left(
x-x_{0}\right) ,$%
\begin{equation*}
\delta _{\sigma }\left( x-x_{0}\right) =C\frac{\chi \left( x\right) }{\left(
2\sqrt{\pi \sigma }\right) ^{3}}\exp \left( -\frac{\left\vert
x-x_{0}\right\vert ^{2}}{4\sigma }\right) ,
\end{equation*}%
\begin{equation}
\int\limits_{G}\delta _{\sigma }\left( x-x_{0}\right) dx=1,  \label{1.20}
\end{equation}%
where the number $C>0$ is chosen such that (\ref{1.20}) holds. The function $%
\delta _{\sigma }\left( x-x_{0}\right) $ approximates the function $\delta
\left( x-x_{0}\right) $ in the distribution sense for sufficiently small
values of $\sigma $. The function $\delta _{\sigma }\left( x-x_{0}\right) $
is acceptable in Physics as a proper replacement of $\delta \left(
x-x_{0}\right) $, since there is no \textquotedblleft true" delta-function
in the physical reality. On the other hand, the above mentioned method of 
\cite{BukhKlib,Kl16} is applicable to the case when $\delta \left(
x-x_{0}\right) $ is replaced with $\delta _{\sigma }\left( x\right) $.
Therefore, it is reasonable from the Physics standpoint to impose condition (%
\ref{1.141}).

\subsection{Main results}

\label{sec:1.1}

Consider the following problem%
\begin{equation}
\Delta u+\frac{k^{2}}{c^{2}\left( x\right) }u=-g\left( x\right) ,x\in 
\mathbb{R}^{3},  \label{1.12}
\end{equation}%
\begin{equation}
\sum\limits_{j=1}^{3}\frac{x_{j}}{\left\vert x\right\vert }\partial
_{x_{j}}u\left( x,k\right) -iku\left( x,k\right) =o\left( 1\right)
,\left\vert x\right\vert \rightarrow \infty .  \label{1.13}
\end{equation}%
We now refer to Theorem 6 of Chapter 9 of the book \cite{V}, Theorem 3.3 of
the paper \cite{V1} as well as to Theorem 6.17 of the book \cite{GT}.
Combining these results with Lemma 1, we obtain that for each $k\in \mathbb{R%
}$ there exists unique solution $u\left( x,k\right) \in C^{6+\alpha }\left( 
\mathbb{R}^{3}\right) ,\forall \alpha \in \left( 0,1\right) $ of the problem
(\ref{1.12}), (\ref{1.13}).

\textbf{Inverse Problem 1 (IP1)}. Suppose that the function $c\left(
x\right) $ satisfying conditions (\ref{1.1})-(\ref{1.3}) is unknown for $%
x\in \Omega $ and known for $x\in \mathbb{R}^{3}\diagdown \Omega .$ Assume
that the following function $f_{1}\left( x,k\right) $ is known 
\begin{equation}
f_{1}\left( x,k\right) =\left\vert u\left( x,k\right) \right\vert ,\forall
x\in S,\forall k\in \left( a,b\right) .  \label{1.14}
\end{equation}%
Determine the function $q\left( x\right) $ for $x\in \Omega .$

\textbf{Theorem 1}. \emph{Consider IP1.} \emph{Let conditions (\ref{1}) and (%
\ref{2}) hold. Let the function }$g\left( x\right) $\emph{\ satisfies
conditions (\ref{1.10})-(\ref{1.141}). Consider two functions }$c_{1}\left(
x\right) ,c_{2}\left( x\right) $\emph{\ satisfying conditions (\ref{1.1})-(%
\ref{1.3}) and such that }$c_{1}\left( x\right) =c_{2}\left( x\right)
=c\left( x\right) $ \emph{for} $x\in \mathbb{R}^{3}\diagdown \Omega .$\emph{%
\ For }$j=1,2$ \emph{let }$u_{j}\left( x,k\right) \in C^{6+\alpha }\left( 
\mathbb{R}^{3}\right) ,\forall \alpha \in \left( 0,1\right) $\emph{\ be the
solution of the problem (\ref{1.12}), (\ref{1.13}) with }$c\left( x\right)
=c_{j}\left( x\right) $\emph{.} \emph{Assume that\ } 
\begin{equation}
\left\vert u_{1}\left( x,k\right) \right\vert =\left\vert u_{2}\left(
x,k\right) \right\vert ,\forall x\in S,\forall k\in \left( a,b\right) .
\label{1.21}
\end{equation}%
\emph{Then }$c_{1}\left( x\right) \equiv c_{2}\left( x\right) .$

IP1 is about the case when the modulus of the total wave field is measured
for $x\in S,k\in \left( a,b\right) $. Consider the function $u_{0}\left(
x,k\right) ,$%
\begin{equation*}
u_{0}\left( x,k\right) =\int\limits_{G}\frac{\exp \left( ik\left\vert x-\xi
\right\vert \right) }{4\pi \left\vert x-\xi \right\vert }g\left( \xi \right)
d\xi .
\end{equation*}%
This function is the solution of the problem (\ref{1.12}), (\ref{1.13}) with 
$c\left( x\right) \equiv 1.$ Since $c\left( x\right) =1$ for $x\in \mathbb{R}%
^{3}\diagdown G,$ then one can consider $u_{0}\left( x,k\right) $ as the
solution of the problem (\ref{1.12}), (\ref{1.13}) for the background
medium. Hence, the function $u_{s}\left( x,k\right) =u\left( x,k\right)
-u_{0}\left( x,k\right) $ can be considered as the wave, which is scattered
due to the inhomogeneous structure of the coefficient $c\left( x\right) $
for $x\in G.$ This is our motivation for posing Inverse Problem 2.

\textbf{Inverse Problem 2 (IP2)}. Suppose that the function $c\left(
x\right) $ satisfying conditions (\ref{1.1})-(\ref{1.3}) is unknown for $%
x\in \Omega $ and known for $x\in \mathbb{R}^{3}\diagdown \Omega .$ Let $%
u_{s}\left( x,k\right) =u\left( x,k\right) -u_{0}\left( x,k\right) .$ Assume
that the following function $f_{2}\left( x,k\right) $ is known 
\begin{equation*}
f_{2}\left( x,k\right) =\left\vert u_{s}\left( x,k\right) \right\vert
,\forall x\in S,\forall k\in \left( a,b\right) .
\end{equation*}%
Determine the function $c\left( x\right) $ for $x\in \Omega .$

\textbf{Theorem 2}. \emph{Consider IP2. Assume that }%
\begin{equation}
c^{2}\left( x\right) \neq 1,\forall x\in S.  \label{1.122}
\end{equation}%
\emph{Let all conditions of Theorem 1 hold, except that (\ref{1.11}) is not
imposed.\ In addition, let (\ref{1.21}) be replaced with} 
\begin{equation}
\left\vert u_{s,1}\left( x,k\right) \right\vert =\left\vert u_{s,2}\left(
x,k\right) \right\vert ,\forall x\in S,\forall k\in \left( a,b\right) ,
\label{1.22}
\end{equation}%
\emph{where }$u_{s,j}\left( x,k\right) =u_{j}\left( x,k\right) -u_{0}\left(
x,k\right) ,j=1,2.$\emph{\ Then }$c_{1}\left( x\right) \equiv c_{2}\left(
x\right) .$

\subsection{The main difficulty}

\label{sec:1.2}

We now outline the main difficulty of proofs of Theorems 1,2. Although the
same difficulty was described in \cite{Kr}, we briefly present it here for
reader's convenience. Below for any number $a\in \mathbb{C}$ its complex
conjugate is denoted as $\overline{a}$. For an arbitrary number $\beta >0$
denote $\mathbb{C}^{\beta }=\left\{ k\in \mathbb{C}:\func{Im}k>-\beta
\right\} .$ Also, denote $\mathbb{C}_{+}=\left\{ k\in \mathbb{C}:\func{Im}%
k\geq 0\right\} .$

Consider an arbitrary point $x_{0}\in S.$ By Lemma 2 (section 2) there
exists a number $\beta >0$ such that the function $u\left( x^{\prime
},k\right) $ admits the analytic continuation from the real line $\mathbb{R}$
in the half-plane $\mathbb{C}^{\beta }$. Since $\left\vert u\left(
x_{0},k\right) \right\vert ^{2}=u\left( x_{0},k\right) \overline{u\left(
x_{0},k\right) },\forall k\in \mathbb{C}^{\beta },$ then the function $%
\left\vert u\left( x_{0},k\right) \right\vert ^{2}$ is analytic as the
function of the real variable $k\in \mathbb{R}$. Hence, (\ref{1.14}) implies
that the function $\left\vert u\left( x_{0},k\right) \right\vert $ is known
for all $k\in \mathbb{R}.$ The main difficulty is linked with zeros of the
function $u\left( x_{0},k\right) $ in the upper half-plane $\mathbb{C}%
_{+}\diagdown \mathbb{R}.$ Indeed, let the number $a=a\left( x_{0}\right)
\in \mathbb{C}_{+}\diagdown \mathbb{R}$ be such that $u\left( x_{0},a\right)
=0.$ Consider the function $\widetilde{u}\left( x_{0},k\right) ,$ 
\begin{equation*}
\widetilde{u}\left( x_{0},k\right) =\frac{k-\overline{a}}{k-a}u\left(
x_{0},k\right) .
\end{equation*}%
Since 
\begin{equation*}
\left\vert \frac{k-\overline{a}}{k-a}\right\vert =1,\forall k\in \mathbb{R},
\end{equation*}%
then $\left\vert \widetilde{u}\left( x_{0},k\right) \right\vert =\left\vert
u\left( x_{0},k\right) \right\vert ,\forall k\in \mathbb{R}$. In addition,
the function $\widetilde{u}\left( x_{0},k\right) $ is analytic in $\mathbb{C}%
^{\beta }.$ Therefore, it is necessary in proofs of Theorems 1,2 to use a
linkage between the function $u\left( x,k\right) $ and the differential
operator in (\ref{1.12}).

\subsection{Published results}

\label{sec:1.3}

Phaseless inverse problems have a central importance in those applications
where only the amplitude of the scattered signal can be measured, while the
phase either cannot be measured or can be measured only with a poor
precision. Some examples are specular reflection of neutrons \cite{BM},
x-ray crystallography \cite{Ladd} and astronomical imaging \cite{Fienup2},
also see \cite{Fienup1} for other applied examples.

The first uniqueness result for the phaseless inverse scattering problem for
the 1-d

Schr\"{o}dinger equation $y^{\prime \prime }+k^{2}y-q\left( x\right)
y=0,x\in \mathbb{R}$ was proven in \cite{Kl9}. Next, it was extended in \cite%
{NHS} to the case of the discontinuous impedance. Also, see \cite{AS} for a
relevant result. A survey can be found in \cite{Kl11}.

There is also a reach literature about the reconstruction of a compactly
supported complex valued function from the modulus of its Fourier transform.
Uniqueness results for this problem were proven in \cite{Kl8,Kl12}. The
majority of works about this problem is dedicated to numerical methods, see,
e.g. \cite{Dob,Fienup2,Fienup1,H,S}. Recently regularization algorithms were
developed for a similar, the so-called \textquotedblleft autocorrelation
problem" \cite{DaiLamm,Hoffman}. In addition, numerical methods were
developed for the phaseless inverse problem of the determination of
obstacles \cite{Iv1,Iv2}. Related problems of synthesis were considered in 
\cite{Al1,Al2,Dob}.

In section 2 we formulate Lemmata 2-8. In section 3 we prove Theorem 1.
Theorem 2 is proven in section 4.

\section{Lemmata 2-8}

\label{sec:2}

Consider the following Cauchy problem for the acoustic equation in the time
domain%
\begin{equation}
v_{tt}=c^{2}\left( x\right) \Delta v,x\in \mathbb{R}^{3},t\in \left(
0,\infty \right) ,  \label{2.1}
\end{equation}%
\begin{equation}
v\left( x,0\right) =0,v_{t}\left( x,0\right) =g\left( x\right) .  \label{2.2}
\end{equation}%
For any appropriate function $f\left( t\right) $ such that $f\left( t\right)
=0$ for $t<0$ let $\mathcal{F}\left( f\right) \left( k\right) $ denotes its
Fourier transform,%
\begin{equation*}
\mathcal{F}\left( f\right) \left( k\right) =\dint\limits_{0}^{\infty
}f\left( t\right) e^{ikt}dt,k\in \mathbb{R}.
\end{equation*}

\textbf{Lemma 2}. \emph{Assume that conditions (\ref{1.1})-(\ref{1.10})
hold. Then there exists unique solution of the problem (\ref{2.1}), (\ref%
{2.2}) such that} $v,v_{t}\in C^{4}\left( \mathbb{R}^{3}\times \left(
0,T\right) \right) ,\forall T>0.$ \emph{Also, for any bounded domain} $\Phi
\subset \mathbb{R}^{3}$ \emph{there exist constants }$B=B\left( \Phi
,c,g\right) >0$ \emph{and }$b=b\left( \Phi ,c,g\right) >0$\emph{\ depending
only on }$\Phi ,c$\emph{\ and }$g$\emph{\ such that the following estimates
hold}%
\begin{equation}
\left\vert D_{x,t}^{\gamma }v\left( x,t\right) \right\vert \leq
Be^{-bt},\forall x\in \Phi ,\forall t>0;\left\vert \gamma \right\vert \leq 3.
\label{2.3}
\end{equation}%
\emph{Furthermore},%
\begin{equation}
u\left( x,k\right) =\mathcal{F}\left( v\right) \left( x,k\right) ,\forall
x\in \mathbb{R}^{3},\forall k\in \mathbb{R},  \label{2.4}
\end{equation}%
\emph{where the function} $u\left( x,k\right) \in C^{6+\alpha }\left( 
\mathbb{R}^{3}\right) ,\forall \alpha \in \left( 0,1\right) ,\forall k\in 
\mathbb{R}$ \emph{is the unique solution of the problem (\ref{1.12}), (\ref%
{1.13}). For every point }$x\in \Phi $\emph{\ the function }$u\left(
x,k\right) $\emph{\ admits the analytic continuation with respect to }$k$%
\emph{\ from the real line in the half-plane} $\mathbb{C}^{b}.$

\textbf{Proof}. Existence and uniqueness of the solution $v\in H^{2}\left( 
\mathbb{R}^{3}\times \left( 0,T\right) \right) ,\forall T>0$ of the Cauchy
problem (\ref{2.1}), (\ref{2.2}) follows from corollary 4.2 of chapter 4 of
the book \cite{Lad}. Consider the function $w\left( x,t\right) =v_{t}\left(
x,t\right) .$ Then $w\in H^{1}\left( \mathbb{R}^{3}\times \left( 0,T\right)
\right) ,\forall T>0$ and this function is the weak solution of the
following problem%
\begin{equation*}
w_{tt}=c^{2}\left( x\right) \Delta w,x\in \mathbb{R}^{3},t\in \left(
0,\infty \right) ,
\end{equation*}%
\begin{equation*}
w\left( x,0\right) =g\left( x\right) ,w_{t}\left( x,0\right) =0,
\end{equation*}%
see chapter 4 of \cite{Lad}. Applying again corollary 4.2 of chapter 4 of
the book \cite{Lad}, we obtain that $w\in H^{2}\left( \mathbb{R}^{3}\times
\left( 0,T\right) \right) ,\forall T>0.$ Hence, Theorem 2.2 of \cite{Ktherm}
implies that $w\in C^{4}\left( \mathbb{R}^{3}\times \left( 0,T\right)
\right) ,\forall T>0.$ Since 
\begin{equation*}
v\left( x,t\right) =\dint\limits_{0}^{t}w\left( x,\tau \right) d\tau ,
\end{equation*}%
then the function $v$ has at least the same smoothness as the function $w$.

To prove (\ref{2.3}), we refer to well known results of Vainberg about the
asymptotic behavior of solutions of Cauchy problems for hyperbolic
equations. More precisely, we refer to Lemma 6 in chapter 10 of the book 
\cite{V} as well as to Remark 3 after this lemma. To apply these results, we
need the non-trapping property of geodesic lines generated by the function $%
c\left( x\right) .$ Since Lemma 1 guarantees this property, then (\ref{2.3})
is true.

To prove connection (\ref{2.4}) between the solution of the problem (\ref%
{1.12}), (\ref{1.13}) and the Fourier transform of the function $v\left(
x,t\right) ,$ we again refer to Lemma 1, Theorem 6 of Chapter 9 of the book 
\cite{V}, Theorem 3.3 of the paper \cite{V1} and to Theorem 6.17 of the book 
\cite{GT}. The assertion about the analytic continuation follows from (\ref%
{2.3}) and (\ref{2.4}). $\square $

The integration by parts in the integral (\ref{2.4}) of the Fourier
transform immediately implies Lemma 3.

\textbf{Lemma 3}. \emph{Assume that conditions (\ref{1.1})-(\ref{1.10})
hold. Then the following asymptotic formulae are valid uniformly for }$x\in 
\overline{G}$%
\begin{equation}
u\left( x,k\right) =-\frac{1}{k^{2}}\left[ g\left( x\right) +O\left( \frac{1%
}{k}\right) \right] ,\left\vert k\right\vert \rightarrow \infty ,k\in 
\mathbb{C}_{+},  \label{2.50}
\end{equation}%
\begin{equation}
u_{s}\left( x,k\right) =\frac{1}{k^{4}}\left[ \left( c^{2}\left( x\right)
-1\right) \Delta g\left( x\right) +o\left( 1\right) \right] ,\left\vert
k\right\vert \rightarrow \infty ,k\in \mathbb{C}_{+}.  \label{2.51}
\end{equation}

Lemmata 4,5 follow immediately from Lemma 3.

\textbf{Lemma 4}. \emph{Assume that conditions (\ref{1.1})-(\ref{1.10})
hold. In addition, assume that there exists a point }$x^{\prime }\in 
\overline{G}_{1}$\ \emph{such that }$g\left( x^{\prime }\right) \neq 0.$ 
\emph{Then the function }$u\left( x^{\prime },k\right) $\emph{\ has at most
finite number of zeros in }$\mathbb{C}_{+}.$

\textbf{Lemma 5}. \emph{Assume that conditions (\ref{1.1})-(\ref{1.10})
hold. In addition, assume that there exists a point }$x^{\prime }\in 
\overline{G}_{1}$\emph{\ such that }$\left( c^{2}\left( x^{\prime }\right)
-1\right) \Delta g\left( x^{\prime }\right) \neq 0.$\emph{\ Then the
function }$u_{s}\left( x^{\prime },k\right) $\emph{\ has at most finite
number of zeros in }$\mathbb{C}_{+}.$

Lemma 6 follows immediately from Proposition 4.2 of \cite{Kl11}.

\textbf{Lemma 6.} \emph{Let }$\beta >0$\emph{\ be a number.\ Let the
function }$d\left( k\right) $\emph{\ be analytic in }$\mathbb{C}^{\beta }$%
\emph{\ and does not have zeros in }$\mathbb{C}_{+}.$\emph{\ Assume that } 
\begin{equation*}
d\left( k\right) =\frac{C}{k^{n}}\left[ 1+o\left( 1\right) \right] \exp
\left( ikL\right) ,\left\vert k\right\vert \rightarrow \infty ,k\in \mathbb{C%
}_{+},
\end{equation*}%
\emph{where }$C\in \mathbb{C}$\emph{\ and }$n\emph{,}L\in \mathbb{R}$\emph{\
are some numbers and also }$n\geq 0$\emph{. Then the function }$d\left(
k\right) $\emph{\ can be uniquely determined for }$k\in \mathbb{C}^{\beta }$%
\emph{\ by the values of }$\left\vert d\left( k\right) \right\vert $\emph{\
for }$k\in \mathbb{R}$\emph{. }

Lemma 7 was actually proven in section 1.2, since the analyticity of the
function $\left\vert u\left( x^{\prime },k\right) \right\vert ^{2}$ for $%
k\in \mathbb{R}$ was proven there.

\textbf{Lemma 7}. \emph{Let the function }$d\left( k\right) $\emph{\ be
analytic for all }$k\in \mathbb{R}.$ \emph{Then the function }$\left\vert
d\left( k\right) \right\vert $\emph{\ can be uniquely determined for all }$%
k\in \mathbb{R}$ \emph{by values of }$\left\vert d\left( k\right)
\right\vert $ \emph{\ for }$k\in \left( a,b\right) $\emph{.}

Lemma 8 is one of versions of the well known principle of the finite speed
of propagation for hyperbolic equations. The proof of this lemma follows
immediately from the standard energy estimate of \S 2 in chapter 4 of the
book \cite{Lad}.

\textbf{Lemma 8}. \emph{Let }$c_{1}\left( x\right) $\emph{\ and }$%
c_{2}\left( x\right) $\emph{\ be two functions satisfying conditions (\ref%
{1.1})-(\ref{1.3}). Also, let conditions (\ref{1}), (\ref{2}) and (\ref{1.10}%
) hold. Assume that }$c_{1}\left( x\right) =c_{2}\left( x\right) =c\left(
x\right) $\emph{\ for }$x\in \mathbb{R}^{3}\diagdown \Omega .$\emph{\ For }$%
j=1,2$\emph{\ let }$v_{j}\in C^{4}\left( \mathbb{R}^{3}\times \left(
0,T\right) \right) ,\forall T>0$\emph{\ be the solution of the problem (\ref%
{2.1}), (\ref{2.2}) with }$c\left( x\right) =c_{j}\left( x\right) $\emph{.
Then there exists a sufficiently small number }$\xi =\xi \left(
c,g,\varepsilon \right) >0$\emph{\ such that} 
\begin{equation*}
v_{1}\left( x,t\right) =v_{2}\left( x,t\right) ,\forall x\in S,\forall t\in
\left( 0,\xi \right) .
\end{equation*}

\section{Proof of Theorem 1}

\label{sec:3}

Consider an arbitrary point $x_{0}\in S.$ Denote 
\begin{equation}
q_{1}\left( k\right) =u_{1}\left( x_{0},k\right) ,q_{2}\left( k\right)
=u_{2}\left( x_{0},k\right) .  \label{3}
\end{equation}%
By Lemma 2 there exists a number $\theta >0$ such that each of functions $%
q_{1}\left( k\right) $ and $q_{2}\left( k\right) $ admits the analytic
continuation in the half-plane $\mathbb{C}^{\theta }.$ It follows from (\ref%
{1.21}) and (\ref{3}) that $\left\vert q_{1}\left( k\right) \right\vert
=\left\vert q_{2}\left( k\right) \right\vert ,\forall k\in \left( a,b\right)
.$ Hence, using Lemma 7, we obtain 
\begin{equation}
\left\vert q_{1}\left( k\right) \right\vert =\left\vert q_{2}\left( k\right)
\right\vert ,\forall k\in \mathbb{R}.  \label{5}
\end{equation}

First, we prove that sets of real zeros of functions $q_{1}\left( k\right) $
and $q_{2}\left( k\right) $ coincide. Let $a\in \mathbb{R}$ be a real zero
of the multiplicity $r_{1}>0$ of the function $q_{1}\left( k\right) .$
Suppose that $a$ is also one of zeros of the function $q_{2}\left( k\right) $
of the multiplicity $r_{2}\geq 0.$ Lemma 4 implies that both numbers $%
r_{1},r_{2}<\infty .$ By (\ref{5})\ 
\begin{equation}
\left\vert \left( k-a\right) ^{r_{1}}\right\vert \cdot \left\vert \widetilde{%
q}_{1}\left( k\right) \right\vert =\left\vert \left( k-a\right)
^{r_{2}}\right\vert \cdot \left\vert \widetilde{q}_{2}\left( k\right)
\right\vert ,\text{ }\forall k\in \mathbb{R},  \label{4.0}
\end{equation}%
where 
\begin{equation}
\widetilde{q}_{1}\left( a\right) \widetilde{q}_{2}\left( a\right) \neq 0.
\label{4.01}
\end{equation}%
Assume, for example that $r_{2}<r_{1}.$ Dividing (\ref{4.0}) by $\left\vert
\left( k-a\right) ^{r_{2}}\right\vert $ and setting $k\rightarrow 0,$ we
obtain $\widetilde{q}_{2}\left( a\right) =0,$ which contradicts to (\ref%
{4.01}). Hence, functions $q_{1}\left( k\right) $ and $q_{2}\left( k\right) $
have the same real zeros.

We now focus on complex zeros in $\mathbb{C}_{+}\diagdown \mathbb{R}$. Since
by Lemma 4 each of functions $q_{1}\left( k\right) ,q_{2}\left( k\right) $
has at most finite number of zeros in $\mathbb{C}_{+},$ then let $\left\{
\eta _{s}\right\} _{s=1}^{n}\subset \left( \mathbb{C}_{+}\diagdown \mathbb{R}%
\right) $ and $\left\{ \sigma _{p}\right\} _{p=1}^{m}\subset \left( \mathbb{C%
}_{+}\diagdown \mathbb{R}\right) $ be zeros of functions $q_{1}\left(
k\right) $ and $q_{2}\left( k\right) $ respectively. Also, let $\left\{
a_{r}\right\} _{r=1}^{m^{\prime }}\subset \mathbb{R}$ be real zeros for both
functions $q_{1}\left( k\right) ,q_{2}\left( k\right) .$ Here each zero is
counted as many times as its multiplicity is.

Consider functions $\widehat{q}_{1}\left( k\right) ,\widehat{q}_{2}\left(
k\right) $ defined as 
\begin{equation}
\widehat{q}_{1}\left( k\right) =q_{1}\left( k\right) \left(
\prod\limits_{s=1}^{n}\frac{k-\overline{\eta }_{s}}{k-\eta _{s}}\right)
\left( \prod\limits_{r=1}^{m^{\prime }}\frac{1}{k-a_{r}}\right) ,k\in 
\mathbb{C}^{\theta },  \label{4.03}
\end{equation}%
\begin{equation}
\widehat{q}_{2}\left( k\right) =q_{2}\left( k\right) \left(
\prod\limits_{p=1}^{m}\frac{k-\overline{\sigma }_{p}}{k-\sigma _{p}}\right)
\left( \prod\limits_{r=1}^{m^{\prime }}\frac{1}{k-a_{r}}\right) ,k\in 
\mathbb{C}^{\theta }.  \label{4.04}
\end{equation}%
Hence, $\widehat{q}_{1}\left( k\right) $ and $\widehat{q}_{2}\left( k\right) 
$ are analytic functions in $\mathbb{C}^{\theta }.$ In addition, it follows
from (\ref{2.50}), (\ref{3}), (\ref{5}), (\ref{4.03}) and (\ref{4.04}) that 
\begin{equation}
\widehat{q}_{j}\left( k\right) =-\frac{1}{k^{m^{\prime }+2}}\left[ g\left(
x_{0}\right) +O\left( \frac{1}{k}\right) \right] ,\left\vert k\right\vert
\rightarrow \infty ,k\in \mathbb{C}_{+},j=1,2,  \label{4.05}
\end{equation}%
\begin{equation}
\widehat{q}_{j}\left( k\right) \neq 0,\forall k\in \mathbb{C}_{+},j=1,2,
\label{4.06}
\end{equation}%
\begin{equation}
\left\vert \widehat{q}_{1}\left( k\right) \right\vert =\left\vert \widehat{q}%
_{2}\left( k\right) \right\vert ,\forall k\in \mathbb{R}.  \label{4.07}
\end{equation}%
Combining (\ref{4.05}), (\ref{4.06}) and (\ref{4.07}) with Lemma 6, we
obtain 
\begin{equation*}
\widehat{q}_{1}\left( k\right) =\widehat{q}_{2}\left( k\right) ,\forall k\in 
\mathbb{R}.
\end{equation*}%
Hence, (\ref{4.03}) and (\ref{4.04}) lead to%
\begin{equation}
q_{1}\left( k\right) \left( \prod\limits_{s=1}^{n}\frac{k-\overline{\eta }%
_{s}}{k-\eta _{s}}\right) =q_{2}\left( k\right) \left( \prod\limits_{p=1}^{m}%
\frac{k-\overline{\sigma }_{p}}{k-\sigma _{p}}\right) .  \label{4.071}
\end{equation}%
Or 
\begin{equation*}
q_{1}\left( k\right) \left( \prod\limits_{p=1}^{m}\frac{k-\sigma _{p}}{k-%
\overline{\sigma }_{p}}\right) =q_{2}\left( k\right) \left(
\prod\limits_{s=1}^{n}\frac{k-\eta _{s}}{k-\overline{\eta }_{s}}\right) .
\end{equation*}%
Or%
\begin{equation}
q_{1}\left( k\right) Y_{1}\left( k\right) +q_{1}\left( k\right) =q_{2}\left(
k\right) Y_{2}\left( k\right) +q_{2}\left( k\right) ,  \label{4.08}
\end{equation}%
where%
\begin{equation}
Y_{1}\left( k\right) =\prod\limits_{p=1}^{m}\frac{k-\sigma _{p}}{k-\overline{%
\sigma }_{p}}-1,  \label{4.09}
\end{equation}%
\begin{equation}
Y_{2}\left( k\right) =\prod\limits_{s=1}^{n}\frac{k-\eta _{s}}{k-\overline{%
\eta }_{s}}-1.  \label{4.010}
\end{equation}

We now calculate the inverse Fourier transform $\mathcal{F}^{-1}$ of
functions $Y_{1}\left( k\right) ,Y_{2}\left( k\right) .$ It follows from (%
\ref{4.09}) that the function $Y_{1}\left( k\right) $ can be represented as%
\begin{equation*}
Y_{1}\left( k\right) =P_{1}\left( k\right) \prod\limits_{p=1}^{m}\frac{1}{k-%
\overline{\sigma }_{p}},
\end{equation*}%
where $P_{1}\left( k\right) $ is a polynomial of the degree less than $m$.
Using the partial fraction expansion, we obtain%
\begin{equation*}
Y_{1}\left( k\right) =\dsum\limits_{j=1}^{\widetilde{m}}\frac{C_{j}}{\left(
k-\overline{\sigma }_{j}\right) ^{r_{j}}},
\end{equation*}%
where $C_{j}\in \mathbb{C}$ are certain numbers, $r_{j}\geq 1$ are some
integers and $\sigma _{j_{1}}\neq \sigma _{j_{2}}$ if $j_{1}\neq j_{2}.$ The
straightforward calculation shows that 
\begin{equation*}
\frac{1}{\left( k-\overline{\sigma }_{j}\right) ^{r_{j}}}=B_{j}\dint%
\limits_{0}^{\infty }t^{r_{j}-1}\exp \left( -i\overline{\sigma }_{j}t\right)
\exp \left( ikt\right) dt,
\end{equation*}%
where $B_{j}\in \mathbb{C}$ is a certain number. Hence, 
\begin{equation}
\mathcal{F}^{-1}\left( Y_{1}\right) :=y_{1}\left( t\right) =H\left( t\right)
\dsum\limits_{j=1}^{\widetilde{m}}K_{j}^{\left( 1\right) }t^{r_{j}-1}\exp
\left( -i\overline{\sigma }_{j}t\right) ,  \label{4.011}
\end{equation}%
where $K_{j}^{\left( 1\right) }\in \mathbb{C}$ are certain numbers and $%
H\left( t\right) $ is the Heaviside function,%
\begin{equation*}
H\left( t\right) =\left\{ 
\begin{array}{c}
1,\text{ if }t>0, \\ 
0,\text{ if }t<0.%
\end{array}%
\right. 
\end{equation*}%
Similarly, using (\ref{4.010}), we obtain 
\begin{equation}
\mathcal{F}^{-1}\left( Y_{2}\right) :=y_{2}\left( t\right) =H\left( t\right)
\dsum\limits_{j=1}^{\widetilde{n}}K_{j}^{\left( 2\right) }t^{r_{j}-1}\exp
\left( -i\overline{\eta }_{j}t\right)   \label{4.012}
\end{equation}%
with certain numbers $K_{j}^{\left( 2\right) }\in \mathbb{C}.$

Next, we apply the operator $\mathcal{F}^{-1}$ to both sides of (\ref{4.08}%
). Using (\ref{2.4}), (\ref{3}), (\ref{4.08}), (\ref{4.011}), (\ref{4.012})
and the convolution theorem, we obtain%
\begin{equation}
v_{1}\left( x_{0},t\right) +\dint\limits_{0}^{t}v_{1}\left( x_{0},t-\tau
\right) y_{1}\left( \tau \right) d\tau =v_{2}\left( x_{0},t\right)
+\dint\limits_{0}^{t}v_{2}\left( x_{0},t-\tau \right) y_{2}\left( \tau
\right) d\tau ,t>0.  \label{4.0151}
\end{equation}%
Denote%
\begin{equation}
y\left( \tau \right) =y_{1}\left( \tau \right) -y_{2}\left( \tau \right) .
\label{4.015}
\end{equation}%
By Lemma 8 $v_{1}\left( x_{0},t\right) =v_{2}\left( x_{0},t\right) :=h\left(
x_{0},t\right) $ for $t\in \left( 0,\xi \right) .$ Hence, (\ref{4.0151}) and
(\ref{4.015}) imply that 
\begin{equation}
\dint\limits_{0}^{t}h\left( x_{0},t-\tau \right) y\left( \tau \right) d\tau
=0,t\in \left( 0,\xi \right) .  \label{4.014}
\end{equation}%
Differentiating equality (\ref{4.014}) twice with respect to $t$ and using (%
\ref{1.11}) and (\ref{2.2}), we obtain%
\begin{equation}
y\left( t\right) +\frac{1}{g\left( x_{0}\right) }\dint\limits_{0}^{t}h_{tt}%
\left( x_{0},t-\tau \right) y\left( \tau \right) d\tau =0,t\in \left( 0,\xi
\right) .  \label{6}
\end{equation}%
This is a homogeneous Volterra integral equation of the second kind. Hence, 
\begin{equation}
y\left( t\right) =0,t\in \left( 0,\xi \right) .  \label{4.016}
\end{equation}

It follows from (\ref{4.011}), (\ref{4.012}) and (\ref{4.015}) that the
function $y\left( t\right) $ is analytic for $t>0$ as the function of real
variable. Hence, (\ref{4.016}) implies that $y\left( t\right) =0$,$\forall
t>0.$ Hence, by (\ref{4.015}) $y_{1}\left( t\right) =y_{2}\left( t\right)
,\forall t>0.$ Therefore, functions $q_{1}\left( k\right) $ and $q_{2}\left(
k\right) $ have the same sets of zeros in $\mathbb{C}_{+}\diagdown \mathbb{R}%
,$ i.e. $\left\{ \eta _{s}\right\} _{s=1}^{n}=\left\{ \sigma _{p}\right\}
_{p=1}^{m}.$ Thus, (\ref{4.071}) implies that 
\begin{equation*}
q_{1}\left( k\right) =q_{2}\left( k\right) ,\forall k\in \mathbb{R}.
\end{equation*}
Therefore, (\ref{2.4}) and (\ref{3}) imply that 
\begin{equation*}
v_{1}\left( x_{0},t\right) =v_{2}\left( x_{0},t\right) ,\forall t>0.
\end{equation*}
Denote $S_{\infty }=S\times \left( 0,\infty \right) .$ Since $x_{0}\in S$ is
an arbitrary point, then 
\begin{equation}
v_{1}\left( x,t\right) =v_{2}\left( x,t\right) :=p\left( x,t\right) ,\forall
\left( x,t\right) \in S_{\infty }.  \label{4.017}
\end{equation}

Hence, it follows from (\ref{2.1}) and (\ref{2.2}) that both functions $%
v_{1},v_{2}$ are solutions of the following initial boundary value problem
outside of the domain $G_{1}$%
\begin{equation*}
\partial _{t}^{2}v_{j}=c^{2}\left( x\right) \Delta v_{j},x\in \mathbb{R}%
^{3}\diagdown G_{1},t\in \left( 0,\infty \right) ,j=1,2,
\end{equation*}%
\begin{equation*}
v_{j}\left( x,0\right) =0,\partial _{t}v_{j}\left( x,0\right) =g\left(
x\right) ,x\in \mathbb{R}^{3}\diagdown G_{1},
\end{equation*}%
\begin{equation*}
v_{j}\mid _{S_{\infty }}=p\left( x,t\right) .
\end{equation*}%
Hence, $v_{1}\left( x,t\right) =v_{2}\left( x,t\right) $ for $x\in \mathbb{R}%
^{3}\diagdown G_{1},t\in \left( 0,\infty \right) .$ Let 
\begin{equation}
\widetilde{p}\left( x,t\right) =\partial _{\nu }v_{1}\left( x,t\right) \mid
_{S_{\infty }}=\partial _{\nu }v_{2}\left( x,t\right) \mid _{S_{\infty }},
\label{4.018}
\end{equation}%
where $\nu =\nu \left( x\right) $ is the unit normal vector at the point $%
x\in S$, which points outside of the domain $G_{1}.$ Hence, using (\ref{2.1}%
), (\ref{2.2}), (\ref{4.017}) and (\ref{4.018}), we obtain inside of the
domain $G_{1}$%
\begin{equation}
\partial _{t}^{2}v_{j}=c_{j}^{2}\left( x\right) \Delta v_{j},x\in G_{1},t\in
\left( 0,\infty \right) ,j=1,2,  \label{4.019}
\end{equation}%
\begin{equation}
v_{j}\left( x,0\right) =0,\partial _{t}v_{j}\left( x,0\right) =g\left(
x\right) ,x\in G_{1},  \label{4.020}
\end{equation}%
\begin{equation}
v_{j}\mid _{S_{\infty }}=p\left( x,t\right) ,\partial _{\nu }v_{j}\left(
x,t\right) \mid _{S_{\infty }}=\widetilde{p}\left( x,t\right) .
\label{4.021}
\end{equation}%
By Lemma 2 $v_{j}\in C^{4}\left( \overline{G}_{1}\times \left[ 0,T\right]
\right) ,\forall T>0.$ In addition, condition (\ref{1.3}) guarantees the
validity of the Carleman estimate for the operator $\partial
_{t}^{2}-c^{2}\left( x\right) \Delta ,$ see Theorem 2.6 in \cite{Kl16}.
Thus, it follows from Theorem 3.1 of \cite{Kl16} that conditions (\ref{1.141}%
), (\ref{4.019}), (\ref{4.020}) and (\ref{4.021}) imply that $c_{1}\left(
x\right) =c_{2}\left( x\right) $ in $G_{1}.$ Finally, since one of
conditions of this theorem is that $c_{1}\left( x\right) =c_{2}\left(
x\right) $ for $x\in \mathbb{R}^{3}\diagdown \Omega ,$ then $c_{1}\left(
x\right) \equiv c_{2}\left( x\right) .$ $\square $

Note that Theorem 3.1 of \cite{Kl16} can also be applied in the case when $%
t\in \left( 0,\infty \right) $ in (\ref{4.019}) and (\ref{4.021}) is
replaced with $t\in \left( 0,T\right) $ for a certain finite number $T>0.$
The proof of Corollary 1 follows immediately from the proof of Theorem 1:
the part, which is before (\ref{4.017}).

\textbf{Corollary 1}. \emph{Let conditions (\ref{1}) and (\ref{2}) hold. Let
the function }$g\left( x\right) $\emph{\ satisfies conditions (\ref{1.10}).
Let }$x_{0}\in S$ \emph{be an arbitrary point. Assume that }$g\left(
x_{0}\right) \neq 0.$\emph{\ Consider two functions }$c_{1}\left( x\right)
,c_{2}\left( x\right) $\emph{\ satisfying conditions (\ref{1.1})-(\ref{1.3})
and such that }$c_{1}\left( x\right) =c_{2}\left( x\right) =c\left( x\right) 
$ \emph{for} $x\in \mathbb{R}^{3}\diagdown \Omega .$\emph{\ For }$j=1,2$ 
\emph{let }$u_{j}\left( x,k\right) \in C^{6+\alpha }\left( \mathbb{R}%
^{3}\right) ,\forall \alpha \in \left( 0,1\right) $\emph{\ be the solution
of the problem (\ref{1.12}), (\ref{1.13}) with }$c\left( x\right)
=c_{j}\left( x\right) $\emph{.} \emph{Assume that}%
\begin{equation*}
\left\vert u_{1}\left( x_{0},k\right) \right\vert =\left\vert u_{2}\left(
x_{0},k\right) \right\vert ,\forall k\in \left( a,b\right) .\emph{\ }
\end{equation*}%
\emph{\ Then }$u_{1}\left( x_{0},k\right) =u_{1}\left( x_{0},k\right)
,\forall k\in \mathbb{R}.$

\section{Proof of Theorem 2}

\label{sec:4}

Let the function $v_{0}\left( x,t\right) \ $be the solution of the Cauchy
problem (\ref{2.1}), (\ref{2.2}) with $c\left( x\right) \equiv 1.$ Denote $%
v_{s}\left( x,t\right) =v\left( x,t\right) -v_{0}\left( x,t\right) .$ Let $%
x_{0}\in S$ be an arbitrary point of the surface $S$. By Lemma 8%
\begin{equation}
v_{s,1}\left( x_{0},t\right) =v_{s,2}\left( x_{0},t\right) :=h_{s}\left(
x_{0},t\right) ,\forall t\in \left( 0,\xi \right) ,  \label{5.1}
\end{equation}%
where $v_{s,j}\left( x,t\right) $ is the function $v_{s}\left( x,t\right) $
for the case when $c\left( x\right) =c_{j}\left( x\right) ,j=1,2.$ It
follows from (\ref{1.141}), (\ref{1.122}) and (\ref{2.51}) that we can apply
the same technique as the one in section 3 before (\ref{4.014}). Hence, (\ref%
{4.014}) is replaced now with%
\begin{equation}
\dint\limits_{0}^{t}h_{s}\left( x_{0},t-\tau \right) y\left( \tau \right)
d\tau =0,t\in \left( 0,\xi \right) .  \label{5.2}
\end{equation}%
By (\ref{2.1}) and (\ref{2.2}) $\partial _{t}^{l}v_{s}\left( x,0\right)
=0,l=0,1,2$ and also 
\begin{equation}
\text{ }\partial _{t}^{3}v_{s}\left( x,0\right) =\left( c^{2}\left( x\right)
-1\right) \Delta g\left( x\right) :=\widetilde{g}\left( x\right) .
\label{5.3}
\end{equation}%
By (\ref{1.141}), (\ref{1.122}) and (\ref{5.3})%
\begin{equation}
\widetilde{g}\left( x\right) \neq 0,\forall x\in S.  \label{5.4}
\end{equation}%
Differentiate equality (\ref{5.2}) three times and use (\ref{5.1}), (\ref%
{5.3}) and (\ref{5.4}). We obtain the following integral equation of the
Volterra type%
\begin{equation*}
y\left( t\right) +\frac{1}{\widetilde{g}\left( x_{0}\right) }%
\dint\limits_{0}^{t}\partial _{t}^{3}h_{s}\left( x_{0},t-\tau \right)
y\left( \tau \right) d\tau =0,t\in \left( 0,\xi \right) .
\end{equation*}%
The rest of the proof is the same as the one in section 3 after (\ref{6}). $%
\square $

\textbf{Corollary 2}. \emph{Let conditions (\ref{1}) and (\ref{2}) hold. Let
the function }$g\left( x\right) $\emph{\ satisfies conditions (\ref{1.10}).
Let }$x_{0}\in S$ \emph{be an arbitrary point. Assume that} $\left(
c^{2}\left( x_{0}\right) -1\right) \Delta g\left( x_{0}\right) \neq 0.$ 
\emph{Consider two functions }$c_{1}\left( x\right) ,c_{2}\left( x\right) $%
\emph{\ satisfying conditions (\ref{1.1})-(\ref{1.3}) and such that }$%
c_{1}\left( x\right) =c_{2}\left( x\right) =c\left( x\right) $ \emph{for} $%
x\in \mathbb{R}^{3}\diagdown \Omega .$\emph{\ For }$j=1,2$ \emph{let }$%
u_{j}\left( x,k\right) \in C^{6+\alpha }\left( \mathbb{R}^{3}\right)
,\forall \alpha \in \left( 0,1\right) $\emph{\ be the solution of the
problem (\ref{1.12}), (\ref{1.13}) with }$c\left( x\right) =c_{j}\left(
x\right) $\emph{.} \emph{Assume that}%
\begin{equation*}
\left\vert u_{1}\left( x_{0},k\right) \right\vert =\left\vert u_{2}\left(
x_{0},k\right) \right\vert ,\forall x\in S,\forall k\in \left( a,b\right) .%
\emph{\ }
\end{equation*}%
\emph{\ Then }$u_{1}\left( x_{0},k\right) =u_{1}\left( x_{0},k\right)
,\forall k\in \mathbb{R}.$

The proof of Corollary 2 follows immediately from the proof of Theorem 2.

\begin{center}
\textbf{Acknowledgments}
\end{center}

This research was supported by US Army Research Laboratory and US Army
Research Office grant W911NF-11-1-0399. The author is grateful to Professors
Paul E. Sacks and Boris R. Vainberg for valuable discussions.

\end{document}